\documentclass[epj]{svjour}
\usepackage{graphics}
\usepackage{graphicx}
\usepackage{epstopdf}
\usepackage{amssymb}   
\usepackage{url}
\usepackage{amsmath}
\begin{document}
\title{Hybrid in-beam  PET- and Compton prompt-gamma imaging aimed at enhanced proton-range verification}

\author{J. Balibrea-Correa\inst{1}
        \and
        J. Lerendegui-Marco\inst{1} 
        \and 
        I. Ladarescu\inst{1}
        \and 
        C. Guerrero\inst{2,3}
        \and
        T. Rodríguez-González\inst{2,3}
        \and
        M. C. Jiménez-Ramos\inst{3,4}
        \and
        B. Fernández-Martínez\inst{2,3}
        \and    
        C. Domingo-Pardo\inst{1}
}
%
\offprints{}          
\institute{Instituto de F{\'\i}sica Corpuscular, CSIC-University of Valencia, Valencia, Spain 
           \and
University of Seville, Seville, 41012 Spain 
           \and
Centro Nacional de Aceleradores (U. Sevilla, CSIC, Junta de Andaluc\'ia), Sevilla, 41092 Spain.
            \and
Department of Applied Physics II, ETSA, University of Seville, Seville, Spain
}

\date{Received: \today / Revised version: \today}

\abstract{
We report on an hybrid in-beam PET and prompt-gamma Compton imaging system aimed at quasi real-time ion-range verification in proton-therapy treatments.  Proof-of-concept experiments were carried out at the radiobiology beam line of the CNA cyclotron facility using a set of two synchronous Compton imagers and different target materials. The time structure of the 18~MeV proton beam was shaped with a series of beam-on and beam-off intervals, thereby mimicking a pulsed proton beam on a long time scale. During beam-on intervals, Compton imaging was performed utilizing the high energy $\gamma$-rays promptly emitted from the nuclear reactions occurring in the targets. In the course of the beam-off intervals in-situ positron-emission tomography was accomplished with the same imagers using the $\beta^{+}$ decay of activated nuclei. The targets used were stacks of different materials covering also various proton ranges and energies. A systematic study on the performance of these two complementary imaging techniques is reported and the experimental results interpreted on the basis of Monte Carlo calculations. The results demonstrate the possibility to combine both imaging techniques in a concomitant way, where high-efficiency prompt-gamma imaging is complemented with the high spatial accuracy of PET. Empowered by these results we suggest that a pulsed beam with a suitable duty cycle, in conjunction with in-situ Compton- and PET-imaging may help to attain complementary information and quasi real-time range monitoring with high accuracy.
\PACS{
      {Prompt-Gamma imaging} {} \and
      {Compton-imaging} {} \and {PET-imaging} {} \and {Ion range verification} {}
     } 
} 
\maketitle
\section{Introduction}
\label{sec:1}
Accurate ion-range determination is a key aspect in modern proton-therapy treatments~\cite{Knopf:13}. This technique allows one to target very precisely the tumor area thanks to the large energy deposition at the end of the proton track (Bragg peak). As a consequence, proton-therapy minimizes damage in neighbouring tissues, thereby reducing also long term secondary effects for such treatments. Hence, it is particularly well-suited for many pediatric cases and tumors close to sensitive organs~\cite{Knopf:13}. However, the full potential of proton therapy is still hindered by the lack of high-accuracy real-time range verification, which would enable to use particle beams as a precise and non-invasive scalpel. This could extend the applicability of proton therapy also to diseases such as ventricular tachycardia and many other cardiovascular disorders~\cite{Durante:2016,Durante:2017,Durante:2019}, thereby enlarging the number of patients benefiting from therapeutic high-energy ion beams.  

As the incident proton beam slows down and stops inside the patient tissues, nuclear reactions take place all along the projectile path producing quasi-instantaneously (prompt) emission of secondary radiation, mainly $\gamma$-rays with energies spanning up to 5-6~MeV~\cite{Krimmer:18}. Thus, from the physics point of view, this radiation is especially well suited to monitor the range of the ion beam because of the high spatial and prompt temporal correlation with the primary proton range. 
On the other hand, from an experimental standpoint, a major challenging aspect for real-time prompt-gamma imaging (PGI) is the small signal-to-background ratio in the clinical environments. The latter is commonly constrained by the low efficiency of the radiation detectors at high $\gamma$-ray energies~\cite{Smeets:2012,Perali:2014,Polf:2015,Munoz:2021}, by the limited detector performance at high counting rates~\cite{Knopf:13} and by the contaminant radiation arising from proton-beam interactions along different parts of the accelerator gantry~\cite{Ortega:2015}, which includes also a dominant contribution from neutrons~\cite{Knopf:13,Min:06,Testa:08,Verburg:2014}. 

Alternatively, ion-range monitoring via in-beam Positron-Emission Tomography (PET) has been extensively applied in clinical conditions~\cite{Parodi:04,Enghardt:04,Miyatake:10} and recent experiments with proton- and helium-beams~\cite{Buitenhuis:2017,Ozoemelam:2020a,Ozoemelam:2020b} have shown very promising results. This methodology is based on the simultaneous detection of two 511~keV annihilation $\gamma$-rays coming from the $\beta^{+}$ decay of short-lived $^{12}$N nuclei (t$_{1/2}$=11~ms) activated by the hadronic interactions of primary ions. Several configurations have been proposed for in-beam PET imaging, from systems based on two detection modules  working in time-coincidence~\cite{Buitenhuis:2017,Ozoemelam:2020a} to complex cylindrical ring configurations~\cite{Tashima:2012,Tashima:2016,Tashima:2020,Yoshida:2020}.
Some of the challenges of the PET approach are related to the relatively long positron range of $^{12}$N (18~mm in water~\cite{Dendooven:2015}), the worsening of resolution along the treatment due to the increasing contribution of longer-lived nuclei~\cite{Ozoemelam:2020a} and the generally large backgrounds at low $\gamma$-ray energies, that lead to very low true statistics for real-time implementation~\cite{Bauer:13,Kurz:15}.



In 2016 the concept of hybrid detection systems was discussed by K.~Parodi~\cite{Parodi:16}, as an alternative to overcome some of the aforementioned limitations. In this new concept, high-energy prompt-gamma imaging is combined with the PET technique. As suggested in Ref.~\cite{Lang:14}, this idea could be implemented by adapting systems based on multiple Compton cameras, originally intended for high-sensitivity three-gamma correlations, which can be exploited for some unstable nuclei, whose $\beta^+$ decay is accompanied by an additional de-excitation $\gamma$-ray~\cite{Lang:14}. As discussed in Ref.~\cite{Parodi:16}, one could expect that hybrid PGI-PET systems will open new perspectives for in-vivo real-time range monitoring. This statement relies on the complementarity of both techniques, as prompt-gamma emission is, on one hand, more promising for real-time monitoring, whereas PET imaging can provide tomographic and functional information, resulting also interesting to monitor physiological processes and tumour response.

In the present work we report on a detection system, originally proposed for nuclear experiments of astrophysical interest~\cite{DOMINGOPARDO:2016}, but which enables the simultaneous implementation of the two aforementioned techniques: in-beam PET and Compton PGI. To the best of our knowledge, this work represents the first implementation and demonstration with ion beams of such an hybrid Compton PGI and PET system, as the one proposed in~\cite{Parodi:16,Lang:14}. 
The performance of the proposed system for in-beam range monitoring via Compton PGI have been studied in detail in a previous work, on the basis of Monte Carlo simulations~\cite{Lerendegui:2022,Lerendegui22a}. Thus, the objective of the present study was twofold. On one side, to perform an experimental validation of the Compton-imaging performance envisaged in our previous MC study~\cite{Lerendegui:2022}. On the other hand, to carry out first proof-of-concept measurements to demonstrate the capability of our hybrid imaging system for performing simultaneously PET and PGI with a proton beam. Furthermore, combining two independent techniques for ion-range monitoring, PET and PGI, may provide an enhanced accuracy from the point of view of the systematic and statistical uncertainties. Finally, realizing both techniques with the same apparatus allows one to perform a comparative study, where the advantages and limitations of combining them within the same hybrid system can be addressed in detail. In Ref.~\cite{Balibrea22a} we presented already preliminary results of one part of these measurements, which are reported here in more detail and in combination with several other results and conclusions.

The measurements were carried out at the 18~MeV Cyclotron facility of the Centro Nacional de Aceleradores (CNA), Seville~\cite{CNA:2020}. Radiobiological research using this type of low energy particle accelerators has attracted significant interest in the last decades concurring with the worldwide expansion of hadron-therapy centers~\cite{BARATTOROLDAN:2020}. Although the proton beam energy used in this work is still between a factor of 5 and 10 lower than the one used in clinical treatments, in combination with the customized set-up described in Sec.~\ref{sec:2}, it still allowed us to carry out a proof-of-concept demonstration of the simultaneous PET-Compton in-beam imaging. Data-reduction, detector calibration and reconstruction algorithms are discussed in Sec.~\ref{sec:3}. Sec.~\ref{sec:4} describes the in-situ calibration measurements made for validating and characterizing the detector performance, both in terms of PET and Compton $\gamma$-ray imaging. The proof-of-concept measurements for the simultaneous PET-Compton imaging are reported in Sec.~\ref{sec:5}. Finally, Sec.~\ref{sec:6} provides a general discussion of the results obtained and future steps.

\section{Experimental setup}\label{sec:2}

Our detection system is based on modular and high efficiency Compton cameras, called i-TED~\cite{DOMINGOPARDO:2016,BABIANO:2020}. They can be operated in synchronous mode, thus enabling both PET- and Compton-imaging at the same time. Although this system has been specifically designed for nuclear physics experiments using the neutron Time-Of-Flight (TOF) technique~\cite{hymns,Babiano:2021}, from the experimental standpoint there are significant similarities with ion-range monitoring in hadron-therapy, which we have described in detail in our previous work~\cite{Lerendegui:2022,Lerendegui22a,Balibrea22a}. In order to maximize detection efficiency each i-TED module consists of two planes of Position Sensitive Detectors (PSD) and each PSD uses largest commercially available LaCl$_{3}$(Ce) monolithic scintillation crystals with a size of 50$\times$50~mm$^2$. Each scintillation crystal is optically coupled to 8$\times$8 pixels Silicon Photomultipliers (SensL ArrayJ-60035-65P-PCB). To enhance further the solid angle for Compton events each absorber plane consists of four, 25~mm thick, PSDs covering an area of 100$\times$100~mm$^2$ in each i-TED module. For a point-like 1~MeV $\gamma$-ray source at 5~cm distance from the front face of the module, the coincidence detection efficiency is $\sim$0.2\%. Regarding neutron-induced backgrounds LaCl$_{3}$(Ce) ensures a small sensitivity to neutron interactions in the detection volume itself~\cite{Lerendegui:2022,Lerendegui22a}, while preserving also a high intrinsic detection efficiency for $\gamma$-rays. Finally, a high time-resolution can be obtained with the implemented acquisition system based on PETsys Front-End Board D version 2 (FEB/D-1024). Coincidence-time resolutions (CTR) of $\sim$500~ps, help to reduce random coincidences and other related backgrounds that could degrade the imaging capabilities of the system. For further details on the i-TED modules and developments the reader is referred to~\cite{BABIANO:2020,BALIBREACORREA:2021} and references therein.

The cyclotron facility at CNA used for the present proof-of-concept experiment consists of a Cyclone 18/9 model equipped with an external beam line for multi-purpose research. The cyclotron accelerates protons and deuterons to 18 and 9 MeV, respectively. The beam is then delivered to the dedicated experimental area through a complex beam extraction system. For further details about the research beam line the reader is referred to~\cite{BARATTOROLDAN:2018}.

During the experiment, a proton beam with an energy of 18~MeV was delivered to the experimental area with current values varying from 500 pA up to 1 nA on target, adding-up a total charge ranging from 212 to 674 nC per irradiation, depending on the total irradiation time and specific duty cycle, which will be explained later.  

\begin{figure}[!htbp]
    \centering
    \includegraphics[width=\columnwidth]{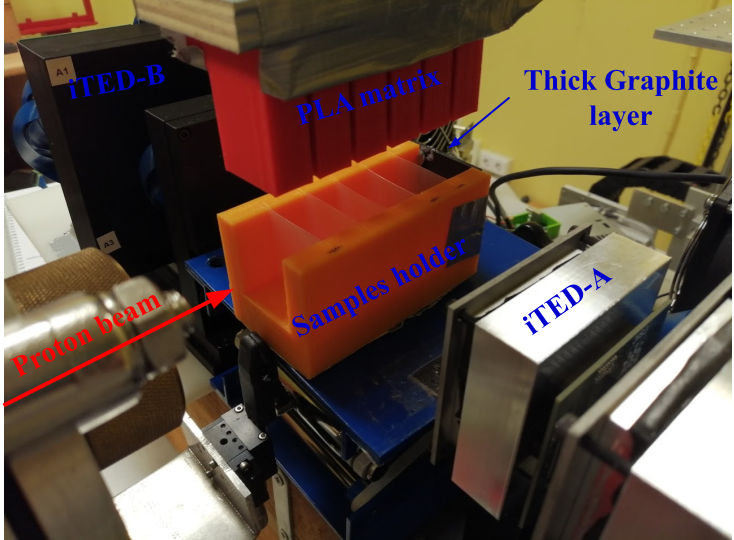}
    \caption{Picture of the setup used in the proof-of-concept experiment at CNA. The proton-beam (red arrow) impinges from the left-hand side and punches through five thin plastic foils surrounded by two i-TED modules in front-to-front configuration.}
    \label{fig:Exp_setup}
\end{figure}

A picture of the experimental setup is displayed in Fig.~\ref{fig:Exp_setup}. Apart from the PET-Compton imaging objective described in this article, this specific configuration was chosen also with the aim of measuring proton-induced $\beta^+$-emitter production cross sections, following a similar methodology as the one reported in the previous works~\cite{Rodriguez-Gonzalez:2020,Rodriguez-Gonzalez:2022}. The proton beam direction is indicated in Fig.~\ref{fig:Exp_setup} by the red arrow. A sample holder (elongated U-shaped piece in the center) was used to simultaneously expose regularly spaced sample-foils to the proton beam. 
The holder was aligned with the proton beam axis, thereby inducing a similar irradiation field for all samples under study. The material of the sample holder was polylactic acid (PLA) plastic with a size of 5.2$\times$5.5$\times$10.3~cm$^{3}$. The central hole had a size of 3.2$\times$4.12~cm$^{2}$. The samples consisted of thin layers of different materials, described below, which were placed in five dedicated slots with a regular gap of 16~mm. The samples had a square size of 41.2$\times$41.2~mm$^{2}$ and a thickness of 0.8~mm. Two different materials were used, Nylon and PMMA, with nominal densities of 1.15 and 1.18~g/cm$^{3}$, respectively.
At the very end of the samples holder a 2~mm thick graphite layer was added, with the twofold purpose of fully stopping the proton beam and registering the proton-current values during the experiment.

\begin{figure}[htbp!]
    \centering
    \begin{tabular}{c c}
    \includegraphics[width=0.48\columnwidth]{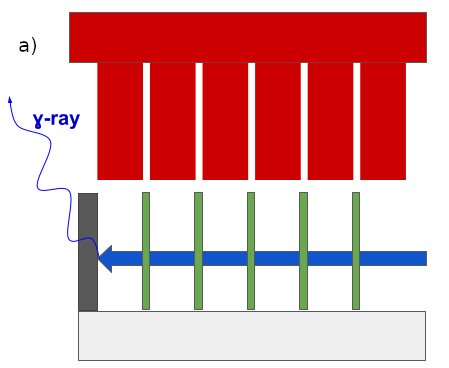} &
    \includegraphics[width=0.48\columnwidth]{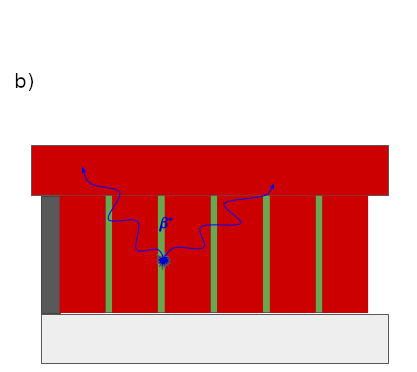}\\
    \end{tabular}
    \caption{Schematic drawing of the experimental setup during the different part of the duty cycles. The thin layers under study are represented by green color, the thick graphite layer by gray and the PLA matrix by red colors, respectively. Panel (a) shows the experimental setup during the beam-on period. In panel (b) is displayed the experimental setup during the beam-off period.}
    \label{fig:experimental_procurement}
\end{figure}

The sample-holder was supplemented with a movable PLA matrix, also shown in Fig~\ref{fig:Exp_setup}. This PLA matrix was remotely inserted and removed. After each proton irradiation the matrix was inserted, thus filling the gap between the samples, acting as $\beta^{+}$ converter and shortening the range of the $e^{+}$ particles. Fig.~\ref{fig:experimental_procurement} shows a schematic drawing of the measurements during proton irradiation (a) and after proton irradiation (b). The measurements made with different materials and beam conditions are described below in Sec.~\ref{sec:5}.

The two i-TED modules used during the experiment were placed front-to-front on both sides of the proton beam axis, fully covering the in-beam PET field of view of the samples under study. In Fig.~\ref{fig:Exp_setup} the i-TED-A module had an aluminum housing and was placed at 5.9~cm from the sample holder. The second one, i-TED-B, had a black PLA encapsulation and was placed at 5.1~cm from the sample holder. The distance between the front face of the scatter plane of both i-TED modules was 27.5~cm. The separation between detection planes in each i-TED was set to only 2.6~cm, thereby favoring coincidence efficiency versus angular resolution~\cite{BABIANO:2020}.

The 10 PSDs embedded in both i-TED modules comprised a total of 640 readout channels, which were synchronously acquired by means of two PETsys Front-End Board D version 2 (FEB/D-1024) modules, synchronized by means of a clock\&trigger module that used LVDS signals at 400~Mbit/s ~\cite{DIFRANCESCO:2016}. The data-stream was read with a Gigabit ethernet connection via the PCI-express board in the acquisition computer for its posterior analysis with the dedicated reconstruction software described in Sec.~\ref{sec:3}. The performance of the i-TED detectors for both Compton and PET imaging is described below in Sec.~\ref{sec:4}.

\section{Data reduction and algorithms}\label{sec:3}

\begin{figure}
    \centering
    \includegraphics[width=\columnwidth]{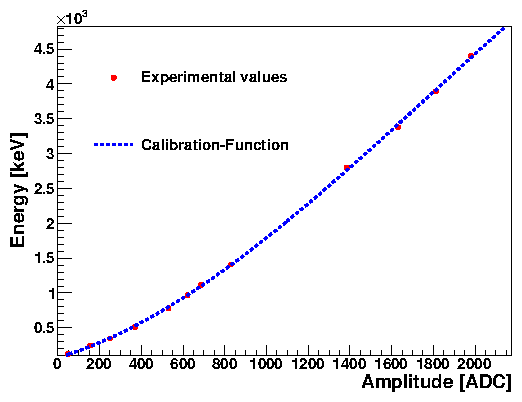}
    \caption{Energy calibration for an individual detector of i-TED-A. The experimental data from the $^{152}$Eu calibration source and $\gamma$-ray lines PG identified during beam-on periods are represented by the red dot points. The fitted third degree polynomial is plotted by the dashed blue line.}
    \label{fig:Energy Calibration}
\end{figure}

The individual detectors of each i-TED module were calibrated in energy using a point-like $^{152}$Eu $\gamma$-ray source. In addition, the $\gamma$-ray lines corresponding to the strongest nuclear reactions identified during the proton irradiations were included in the energy calibration. The latter correspond to the $^{12}$C($p$,$p^{\prime}\gamma$)$^{12}C$ and $^{16}$O($p$,$x\gamma$)$^{12}C$ nuclear reactions, both emitting 4.4 MeV $\gamma$-quanta~\cite{Verburg:2014}. The corresponding first and second escape peaks were also clearly visible and included in the calibration, as well as the 511 keV $\gamma$-ray from the detection of $\beta^{+}$ annihilation events during the beam-on periods. A third polynomial degree was used to model the energy-ADC relation over this large deposited energy range. An example of the individual PSD energy calibration is displayed in Fig.~\ref{fig:Energy Calibration}. The experimental ADC values for the different $\gamma$-ray transitions are represented by red markers, while the fitted polynomial is displayed by the dashed-blue line. 

The energy detection thresholds obtained for i-TED-A and i-TED-B were of 100~keV to 250~keV, respectively. The $\gamma$-ray interaction positions in the individual PSDs were reconstructed following the procedure described in~\cite{BALIBREACORREA:2021}. The absolute $\gamma$-ray position interaction was obtained from the reconstructed intrinsic 3D position coordinates and the location of each PSD, which was well known by set-up construction.

\begin{figure}
    \centering
    \includegraphics[width=\columnwidth]{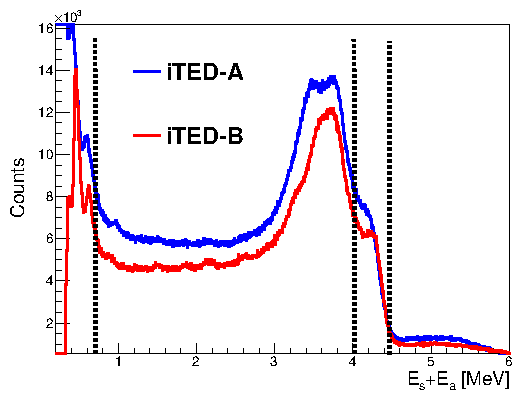}
    \caption{Add-back deposited energy spectrum of i-TED-A (blue) and i-TED-B (red) during the beam-on period. The dashed lines corresponds, at low energy to the energy threshold for imaging, and for high deposited energy the 4.4 MeV $\gamma$-ray line used for Compton imaging.}
    \label{fig:deposited_Energy_Compton}
\end{figure}

The i-TED energy-calibrated add-back time-coincidence spectra between detection planes registered during the beam-on period of configuration (I) explained in the next section is displayed in Fig.~\ref{fig:deposited_Energy_Compton}. The time-coincidence window between different detection planes was set to 10~ns, which allowed to reduce random coincidences with other background sources. The deposited energy spectra registered during the different configurations described later in Sec.~\ref{sec:5} were similar. This result may indicate that the spectra are dominated by the inelastic reactions in carbon nuclei of the thick graphite layer, an effect which is expected from the Bragg curve~\cite{Krimmer:18,Durante:2016} and the large amount of material in the thick graphite target when compared to the thin samples of Nylon and PMMA.

The energy window of 4.0-to-4.6~MeV was chosen for the beam-on periods in order to perform Compton imaging (see high energy dashed lines in Fig.~\ref{fig:deposited_Energy_Compton}). A high energy threshold for the individual absorber PSD detectors, 700 keV, was used for the Compton imaging reconstruction aiming at reduce artifacts in the images due to time-correlated pair production events and random coincidences with 511 keV. The image plane chosen for the Compton reconstruction corresponds to the axial direction of the samples holder, i. e. 8.5~cm from the frontal face of the i-TED-A module. Finally, the Compton images in this work were obtained by implementing the analytical inversion algorithm based on spherical harmonics published by Tomotani and Hisarawa in 2002~\cite{Tomitani:02}. This inversion formula, which is based on an infinite Legendre polynomial expansion, leads to an approximate solution given by a unit vector in the image space, $\vec{s}$. The image at that vector position is described by

\begin{equation}
    f(\vec{s})\approx\int^{cos\omega_{max}}_{cos\omega_{min}}{dcos\omega}\int_{S}{d\vec{t} k^{-1}(\vec{t},\vec{p};cos\omega)g(\vec{t};cos\omega)},
\end{equation}

where $\vec{t}$ is a unit vector into the projection space, $\omega_{min}$ and $\omega_{max}$ are the minimum and maximum Compton scattering angles that can be measured experimentally with the configuration used for the i-TED modules, $g(\vec{t};cos\omega)$ is the projection data in the image space and $k^{-1}(\vec{t},\vec{p};cos(\omega))$ is the inversion kernel. This kernel is defined as
\begin{equation}
        k^{-1}(\vec{t},\vec{p};\omega)=\sum^{N_{max}}_{n=0}{\frac{2n+1}{4\pi H_{n}}P_{n}(cos\omega)P_{n}(\vec{s}\cdot\vec{t})}
\end{equation}

with $H_n$ given by the formula

\begin{equation}
    H_{n}=\int^{cos\omega_{max}}_{cos\omega_{min}}{\sigma(cos\omega)P^{2}_{n}(cos\omega)dcos\omega}.
\end{equation}

In the latter expression $P_{n}$ is the Legendre polynomial of order $n$ and $\sigma(cos\omega)$ is the Klein-Nishina Compton differential cross-section~\cite{Klein:29}. $N_{max}$ is the maximum number of terms involved in the polynomial expansion and it must be chosen according to the angular resolution of the experimental apparatus, i.e., the experimental angular resolution of the i-TED modules that depends on the selected deposited energy window and the distance between detection planes.

The complexity of this algorithm leads to a large computational cost in order to reconstruct a Compton image with sufficient resolution. For this reason, the algorithm was implemented in this work for GPU devices using the CUDA 11.1 toolkit~\cite{CUDA}. This methodology allows for a speed-up factor of about $\sim$121, when compared to the singled-threaded CPU version~\cite{Lerendegui:2022}. Additionally, and aiming at quasi-real time image reconstruction in clinical studies, $H_{n}$ was pre-computed for a wide range of $\gamma$-ray energies and Compton scattering angles corresponding to the detectable angular range between $\omega_{min}$ and $\omega_{max}$. The $H_{n}$ values were saved in a table format for its posterior use, thus enhancing further the speed of the image reconstruction.

\begin{figure}
    \centering
   \includegraphics[width=\columnwidth]{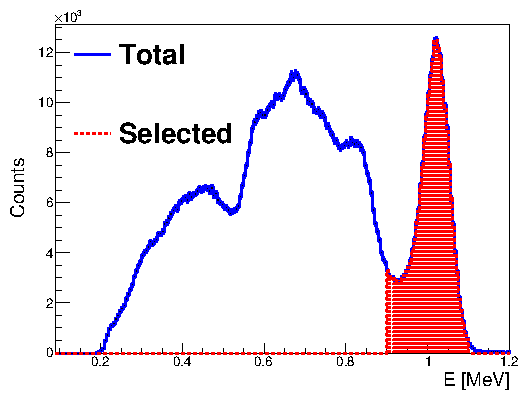}
    \caption{Add-back deposited time-coincidence energy spectrum of the i-TED modules during the beam-off period. The selected deposited energy window for PET imaging is displayed by the shadowed red region.}
    \label{fig:deposited_Energy_PET}
\end{figure}

The add-back energy-calibrated spectrum of the i-TED modules during the beam-off period is displayed in Fig.~\ref{fig:deposited_Energy_PET}. The total add-back spectrum is displayed in blue, while the selected window for the $\beta^{+}$ full-energy coincidence peak used for PET imaging is shown by the red region. This spectrum was obtained using also a time-coincidence window of 10~ns between the individual detectors of different i-TED modules. 

The PET images were reconstructed using a simple analytical algorithm, where straight lines of response (LOR) between the $\gamma$-ray interaction 3D-positions at each i-TED detector were intersected with the central axial plane. The latter PET imaging plane coincides with the one used for Compton imaging during in the beam-on periods, which was at 85~mm from the front face of the i-TED-A module.

\begin{figure}
    \centering
    \includegraphics[width=\columnwidth]{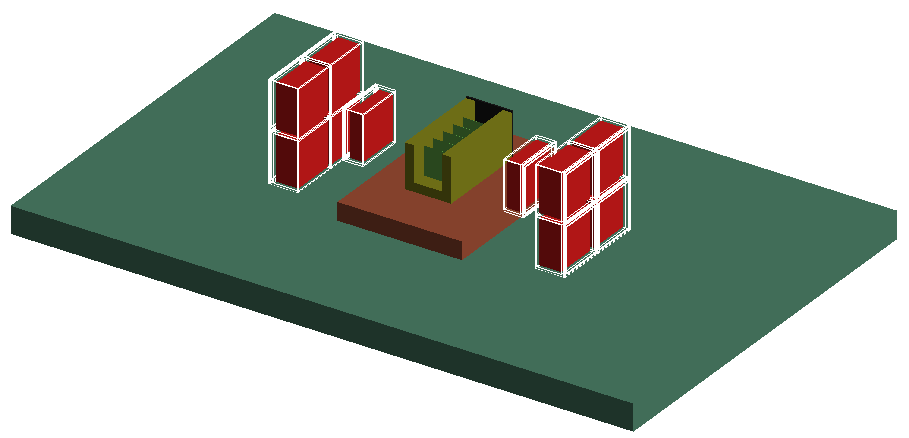}
    \caption{Geometry of the experimental setup as implemented in the MC simulation. See text for details.}
    \label{fig:MC_setup}
\end{figure}

A detailed geometry of the experimental setup was implemented in a C++ Monte Carlo (MC) application based on the \textsc{Geant4} toolkit, version 4.10.6~\cite{ALLISON:2016}. The MC simulation included the standard electromagnetic package option 3, the radioactive decay, and the packages commonly used in hadron-therapy simulations~\cite{Verburg:2012}. These calculations were helpful for the reliable interpretation of the experimental results, as described later in sections~\ref{sec:5} and ~\ref{sec:6}. The geometry implemented in the MC code is displayed in Fig.~\ref{fig:MC_setup}. For Compton imaging it includes the two i-TED modules, the sample holder with the samples and the thick graphite layer. For PET imaging the geometry includes also the PLA converter matrix, not shown in Fig.~\ref{fig:MC_setup}.
In the simulations, the intrinsic resolutions for both position- and energy-response were included according to the laboratory characterization described in the references~\cite{BALIBREACORREA:2021,Olleros:2018}. The energy dependence of the energy resolution for the PSD, $R(E)$, was determined from the energy calibration procedure using a functional of the form

\begin{equation}
    R(E)=\sqrt{a+b/E},
\end{equation}

where the $a$ and $b$ parameters were adjusted from the experimental energy calibrations. In order to perform a realistic comparison the MC-calculated Compton- and PET-images were reconstructed by implementing the same algorithms and conditions applied to the experimental data.

A final consideration is required regarding the effect of the proton-beam divergence. As protons interact with the sample material along the beam path a spreading is introduced in the transversal spatial profile of the beam. Because the Compton- and PET-imaging algorithms used in this work are based on a single 2D image plane for the reconstruction, the images obtained will reflect also the broadening due to beam straggling effects. In order to account for this experimental effect a dedicated simulation of the proton beam passing through the stack of samples was performed for the three different configurations described later in Sec.~\ref{sec:5}. This calculation included the different materials and positions in the experimental setup, as shown in Fig.~\ref{fig:MC_setup}. For each simulation the 3D vertex locations of the inelastic proton scatterings at the different irradiated layers were registered. 
\begin{figure}
    \centering
    \includegraphics[width=\columnwidth]{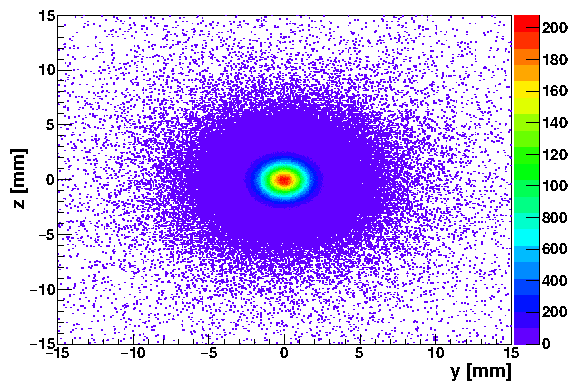}
    \caption{MC simulation of the proton-beam profile registered in a thick graphite layer.}
    \label{fig:Example_beamSpot}
\end{figure}
As an example, Fig.~\ref{fig:Example_beamSpot} shows the proton-beam profile calculated at the graphite layer in the first configuration discussed later in Sec.~\ref{sec:5}. These distributions will be used afterwards as initial emission-vertex distributions for the subsequent MC simulation of the prompt $\gamma$-rays and $e^{+}$ particles in the beam-on and beam-off modes, respectively. The aim for this calculation is twofold, speed-up simulation process getting reliable results at a reasonable time, estimate number of expected inelastic interactions for the individual layers aiming to further interpretation of the comparison between experimental and Monte Carlo results.

\begin{figure}
    \centering
    \includegraphics[width=\columnwidth]{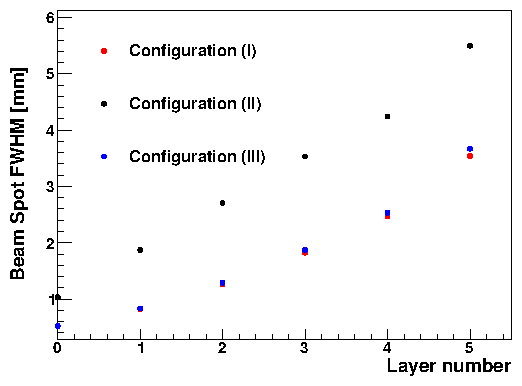}
    \caption{MC-calculated FWHM values for the proton beam spatial distribution at the different layers and for the different configurations.}
    \label{fig:BeamSpotFWHM}
\end{figure}
As expected, the full-width-at-half-maximum (FWHM) values of the proton-beam distributions at the different layers increase along the beam direction, as shown in Fig.~\ref{fig:BeamSpotFWHM}. The main difference between the different configurations (see Sec.~\ref{sec:5}) was obtained for the configuration (II), where a thick 0.8~mm Nylon-based degrader was placed just before the sample holder to reach lower incident proton energies.

Finally, for the proper interpretation of the results shown later in Sec.~\ref{sec:5} it is worth to indicate that for both, PET and Compton, the image reconstruction is made from the reference system of i-TED-B. Thus, the direction of the proton beam in the reconstructed images is from right-to-left, being the left-hand side the position of the thick graphite layer (charge integrator) and the right hand side the entrance of the proton beam, as it is schematically shown in panel (a) of Fig.~\ref{fig:experimental_procurement}.

\section{In-situ characterization of the Compton-PET imaging system}\label{sec:4}

In this section we report on the performance of the two i-TED modules for Compton and PET imaging by comparing MC simulations with data acquired using point-like radioactive sources and well-controlled proton irradiations during the proof-of-concept experiment. The goal of this work is to validate the implemented image-reconstruction algorithms and characterize the systematic behaviour of our detection set-up, as a preceding step to the in-beam Compton-PET imaging application described later in Sec.~\ref{sec:5}.

\subsection{PET imaging}\label{sec:4.1}

The PET performance was experimentally characterized by using a standard point-like $^{22}$Na calibration source placed at the sample-holder slot positions number two, three and four. These positions match with the central region of the imaging system and cover the entire PET field of view between the scatter planes of both i-TED modules. The $^{22}$Na sample was placed at the axial-beam distance, thereby matching well the Compton and PET reconstruction planes. 

\begin{figure*}
\centering
\begin{tabular}{c c c}
   \includegraphics[width=0.7\columnwidth]{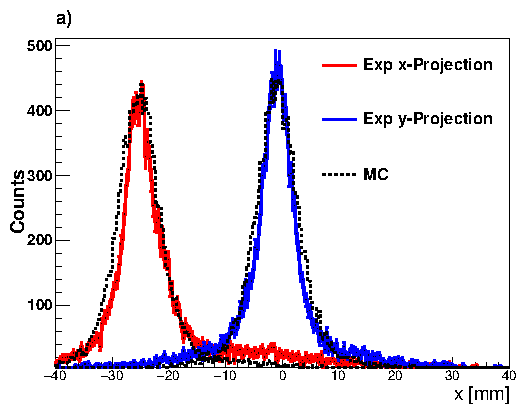} & \includegraphics[width=0.7\columnwidth]{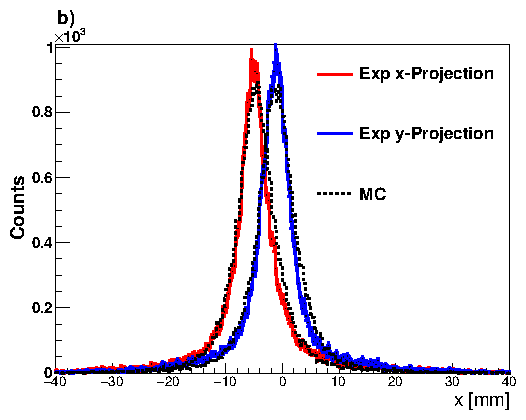} &
   \includegraphics[width=0.7\columnwidth]{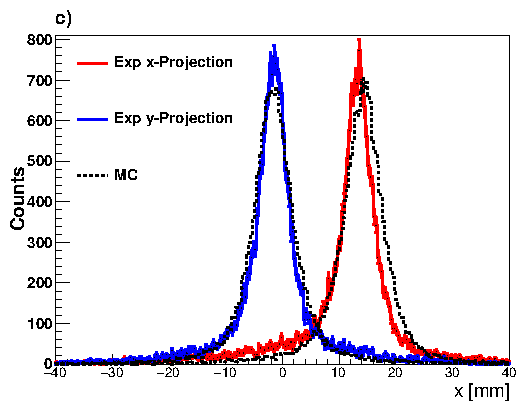}\\
\end{tabular}
\caption{x- (red) and y-axis (blue) projections of the PET images for the $^{22}$Na calibration source at different positions along the sample holder. The result of the MC simulation is shown by the dashed-black curve. Panels (a), (b) and (c) correspond to slots number four, three and two in the sample holder (see also Sec.~\ref{sec:5}).}
\label{fig:Na22_Images}
\end{figure*}

Projections obtained from the PET distributions along the x- (red) and y-axis (blue) for the different source positions are displayed in Fig.~\ref{fig:Na22_Images}. These results were obtained using the simple analytical algorithm described in the previous section. In general, the experimental results are in agreement with the MC simulation, which is shown with a black-dashed line.

For this comparison, the MC simulations were scaled to match the height of the experimental distributions. 
The satisfactory agreement found for peripheral source positions (panels a and c in Fig.~\ref{fig:Na22_Images}) indicates that pin-cushion effects are reasonably well treated by means of the implemented SVM-method described in~\cite{BALIBREACORREA:2021}. The good agreement for both x- and y-axis projections of the central source position (panel b in Fig.~\ref{fig:Na22_Images}), serves to validate also the intrinsic 3D position resolution obtained in our previous work~\cite{BALIBREACORREA:2021} and implemented here in the MC-model. 

\begin{table}[htb]
    \centering
    \begin{tabular}{c|c|c}\hline
    Position & FWHM x [mm] & FWHM y [mm]\\ \hline
    2 & 6.1(1) & 6.2(1) \\
    3 & 5.6(1) &  6.1(1) \\
    4 & 7.4(1) & 6.9(1) \\ \hline
    \end{tabular}
    \caption{Experimental x- and y-axis FWHM calculated from the experimental $^{22}$Na PET image at the different sample positions.}
    \label{tab:Na22}
\end{table}

The PET image resolution displayed in Table~\ref{tab:Na22} for both axis has been determined from the FWHM of a Gaussian fit to the experimental data. As expected, the best resolution in the x-axis is obtained for the central source location (position number three). As one moves away from this point, slightly broader FWHM values are obtained (positions 2 and 4) reflecting a degradation of the PET resolution. This effect can be ascribed to the intrinsic resolution spatial profile of the PSDs, which becomes worse for the peripheral region of the monolithic crystals~\cite{BALIBREACORREA:2021}. 

The average spatial resolution for PET shows an average value of 6.4(6)~mm FWHM. This resolution is in agreement with the one determined by means of MC simulations using individual crystal position reconstructions reported in a previous work~\cite{BALIBREACORREA:2021}, and in line with similar pre-clinical PET-imaging prototypes (5-6~mm) as reported, for example, in~\cite{Parodi:2018} and references therein.

\subsection{Compton imaging}

\begin{figure*}
    \centering
   \begin{tabular}{c c}
   \includegraphics[width=\columnwidth]{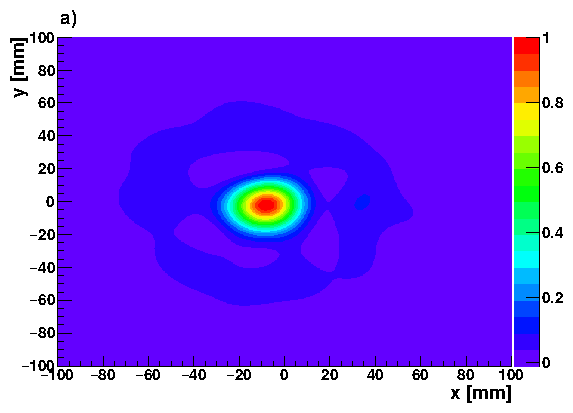} &
   \includegraphics[width=\columnwidth]{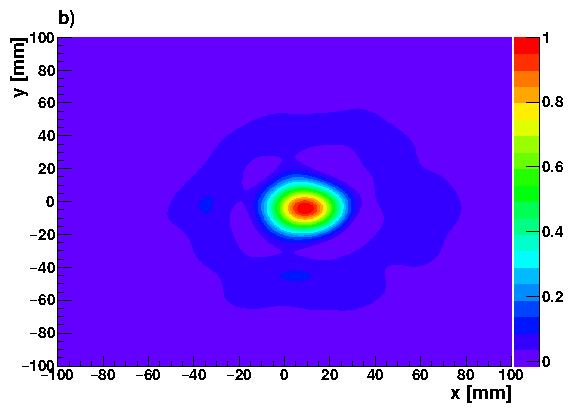} \\
   \includegraphics[width=\columnwidth]{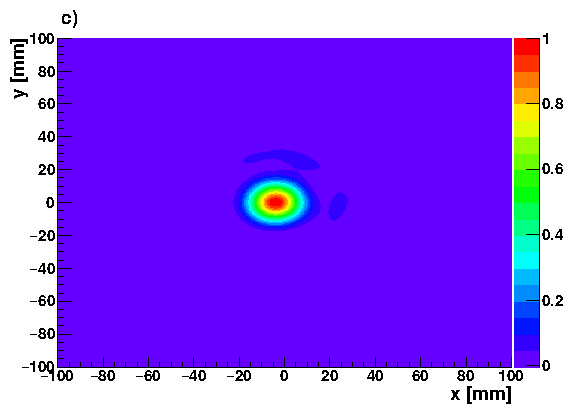} &
   \includegraphics[width=\columnwidth]{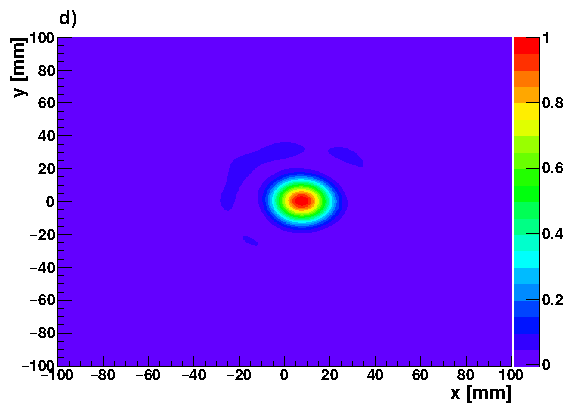} \\
    \end{tabular}
  \caption{Compton images using the 4.4~MeV $\gamma$-ray transition from experimental data (top panels a and b) and Monte Carlo simulations (bottom panels c and d) for the thick graphite layer placed at the third slot position of the sample holder irradiated with the proton beam at CNA. See text for details.}
    \label{fig:i-TED-Run5-Compton}
\end{figure*}

The 2D Compton imaging reconstruction capability and resolution for both i-TED modules at high $\gamma$-ray energy was verified by means of a short dedicated measurement with proton beam, where a 1.8~mm thick graphite layer was placed in the third position of the sample holder, close to the geometrical center of the experimental setup. The incident beam energy was 18~MeV. The 2D images reconstructed with the analytical Compton algorithm described in the previous section are displayed in Fig.~\ref{fig:i-TED-Run5-Compton}. For comparison purposes, the latter figure also shows a point-like 4.4~MeV source simulated at the same position. Differences between experimental and calculated distributions can be ascribed, to some extent, to the spatial profile of the proton beam, of about 2~mm FWHM (Fig.~\ref{fig:BeamSpotFWHM}), not included here in the simulation.


A small systematic shift in the x-position was identified after a detailed comparison between the images reconstructed in panels (a) and (b) of Fig.~\ref{fig:i-TED-Run5-Compton}. Indeed, the position of the graphite layer is reconstructed with a difference of 10(1)~mm between i-TED-A and i-TED-B modules along the x-axis. Since the distance between the detectors was well under control, this shift rather indicates that the central axis of both i-TED modules were not perfectly orthogonal to the beam direction (z-axis). According to the differences in the reconstructed positions and the distance between frontal faces of i-TED modules an angle deviation of about 2-3$^{\circ}$ with respect to the nominal 90$^{\circ}$ orientation has been estimated. No attempt has been made to correct for such systematic bias because it does not compromise the imaging results and conclusions of the present work. On the other hand, a more precise and reliable mechanical structure is in preparation for future similar measurements.

\begin{table}[htb]
    \centering
    \begin{tabular}{c|c|c}\hline
     Detector & FWHM x [mm] & FWHM y [mm]\\ \hline
        i-TED-A & 24.73(5) & 25.12(3)\\
        i-TED-A (MC) & 18.3 & 18.0 \\
        i-TED-B & 24.26(6) & 24.73(4)\\ 
        i-TED-B (MC) & 19.6 & 19.4 \\ \hline
    \end{tabular}
    \caption{Experimental and Monte Carlo FWHM calculated from the Compton images reconstructed using the thick graphite layer in the third position of the samples holder.}
    \label{tab:i-TED-Compton}
\end{table}

The FWHM values obtained for both x- and y-axis projections of the reconstructed Compton images (Fig.~\ref{fig:i-TED-Run5-Compton}) are reported in Tab.~\ref{tab:i-TED-Compton}. As expected, the experimental resolutions obtained for both i-TED modules are rather comparable. On the other hand, a difference of $\sim$5-6~mm is found between simulation and experiment. This is most probably to be ascribed the combination of point-like simulation and the worsening of the intrinsic spatial resolution with increasing $\gamma$-ray energy~\cite{Smeets:2012}, an effect which was not included in the simulation.


\section{Proof-of-concept for hybrid Compton-PET measurements}\label{sec:5}

The proof-of-concept measurements with the proton beam were performed in cycles separated in well defined beam-on and beam-off periods. It is worth to emphasize that the different duty-cycles of the proton beam and the two sample materials were chosen in order to determine production cross-sections of some specific $\beta^{+}$ emitters at different proton-beam energies~\cite{Rodriguez-Gonzalez:2020,Rodriguez-Gonzalez:2022}. For the present work this variety of materials and beam-conditions enabled for a systematic study of both Compton- and PET-techniques in terms of accuracy, sensitivity and repeatability.

During beam-on the $\beta^{+}$ converter matrix was placed off-beam, as schematically shown in Fig.~\ref{fig:experimental_procurement}-(a), thereby avoiding any interfere of the matrix with the incoming proton beam. At this stage all samples under study were irradiated by the proton particles at the same time and in similar amounts. Because of the proton-beam energy loss at each sample, different proton energy ranges were covered at each slot position along the sample holder. The beam was fully stopped in the last layer, a 1.8~mm thick graphite foil, as shown in Fig.~\ref{fig:experimental_procurement}.

In the beam-on intervals PG Compton imaging is exploited by means both i-TED modules, which could provide independent spatial information from the high-energy prompt gamma-rays emitted. In the beam-off periods the PLA matrix was swiftly inserted using a remote mechanical actuator, as shown in Fig~\ref{fig:experimental_procurement}-(b). 
The converter matrix allowed to shorten the mean free-path of the $e^{+}$ particles significantly, thus enabling also a precise PET imaging. 


Three different sample configurations and duty cycles were used. For sake of clarity, in the following sections we refer to those configurations as the following:

\begin{itemize}
    \item Configuration (I): Five Nylon layers and one thick graphite beam stopper at the end.
    \item Configuration (II) Five Nylon layers and, a 0.8~mm thick Nylon based proton-beam energy degrader just before the Nylon samples and one thick graphite beam stopper at the end.
    \item Configuration (III) Five PMMA layers and one thick graphite beam stopper at the end.
\end{itemize}

\begin{figure}[htbp!]
    \centering
    \includegraphics[width=\columnwidth]{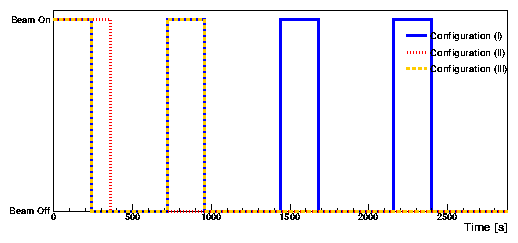}
    \caption{Schematic drawing of the experiment time structure used for the individual configurations used during the experiment. See text for details.}
    \label{fig:time_lapses_configurations}
\end{figure}
The time structure of beam-on and beam-off periods for each individual configuration is graphically represented in Fig.~\ref{fig:time_lapses_configurations}. In configuration (I) there were four beam-on irradiations of 240~s duration, each of them followed by 480~s of beam-off (blue-solid line in Fig.~\ref{fig:time_lapses_configurations}). Configuration (II) consisted in one long irradiation of 360~s followed by a long beam-off (red-dashed line). Configuration (III) consisted of 2 beam-on periods of 240~s duration, followed by 480~s of beam-off (dashed-orange line).

The following sections describe the PET- and Compton imaging results obtained for the three configurations just explained. For each configuration, the experimental results will be compared with the MC simulations. The three configurations are structured into beam-on (PG-Compton) and beam-off (PET) subsections.

\subsection{Results for Configuration (I)}


\subsubsection{Compton imaging in the beam-on lapses}

\begin{figure}[thbp!]
    \centering
   \begin{tabular}{c c}
    \includegraphics[width=0.48\columnwidth]{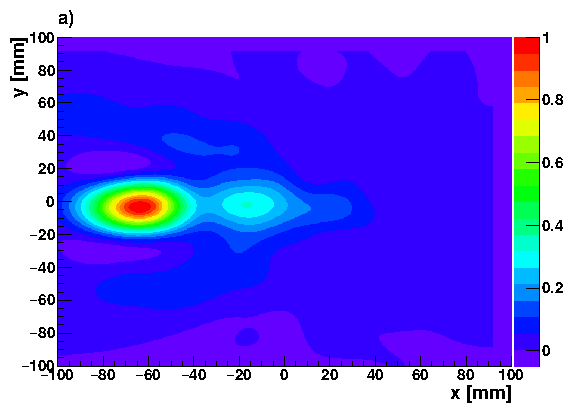} &
    \includegraphics[width=0.48\columnwidth]{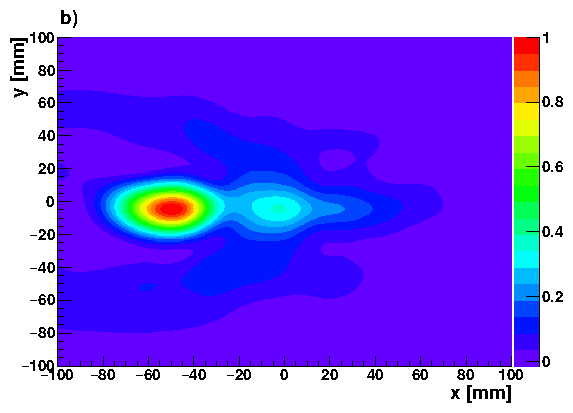} \\
    \includegraphics[width=0.48\columnwidth]{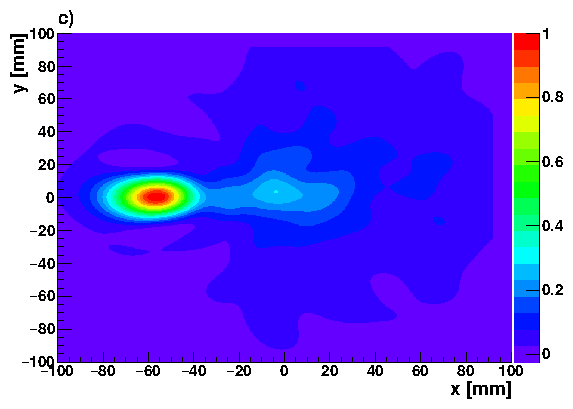} &
    \includegraphics[width=0.48\columnwidth]{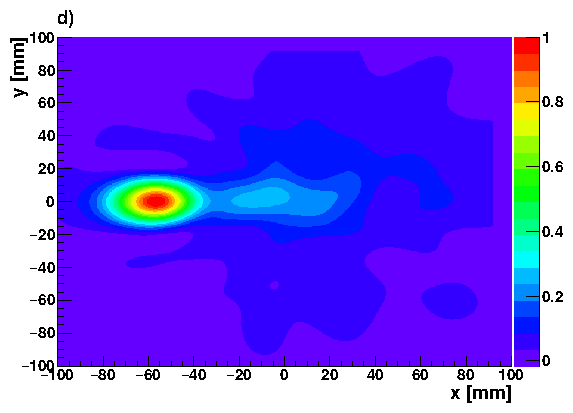} \\
    \end{tabular}
    \caption{Reconstructed experimental (top) and MC-calculated (bottom) Compton images from 4.4~MeV prompt $\gamma$-rays in configuration (I) for i-TED-A (left) and i-TED-B (right).}
    \label{fig:Compton_Run7}
\end{figure}

The Compton images reconstructed from the experimental data acquired with i-TED-A and i-TED-B during the beam-on lapses of configuration (I) are displayed in panels (a) and (b) of Fig~\ref{fig:Compton_Run7}. Panels (c) and (d) show the reconstructed MC images for i-TED-A and i-TED-B, respectively.

The reconstructed experimental and MC images for this configuration show a prominent maximum, which can be ascribed to prompt $\gamma$-rays emitted at the Bragg peak from the thick graphite layer. A secondary maximum can be inferred, particularly at the central region of the experimental image, corresponding to the geometric center of the setup where detection efficiency is largest. 
The fact that only the signature of the central Nylon layer can be appreciated indicates that detection sensitivity for prompt Compton imaging is not enough to resolve the other four thin Nylon layers, as expected from the MC-simulation. 
The ratio between the first and second maximum in the experimental and MC images changes from $\sim$0.45 down to $\sim$0.35, respectively. This difference could be due to discrepancies between the evaluated cross-sections and/or stopping power in the libraries of the MC code and the actual cross-section values. 

\subsubsection{PET imaging in the beam-off lapses}
\begin{figure}[htbp!]
    \includegraphics[width=\columnwidth]{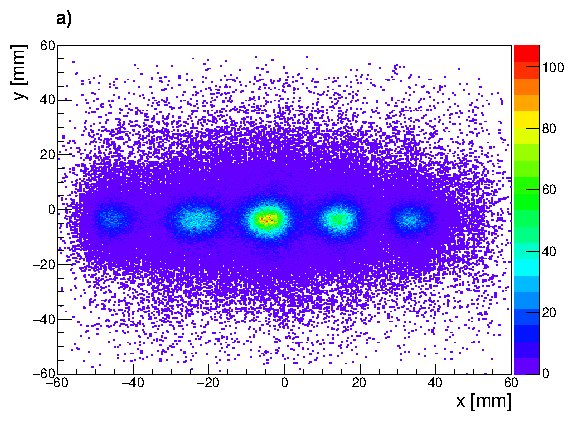} 
    \includegraphics[width=\columnwidth]{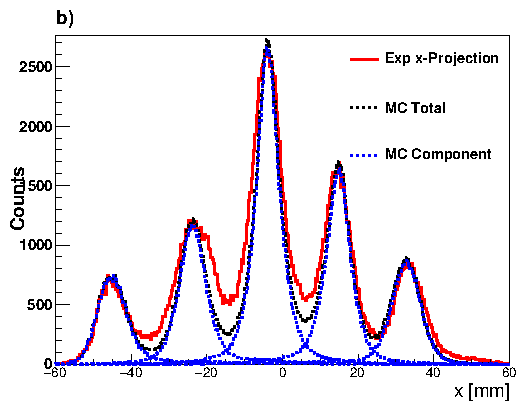} \\
    \caption{Panel (a) shows the experimental 2D PET image for Configuration (I). Panel (b) displays the x-projection (red) together with the MC simulations of the individual contributions (blue) and total (black).}
    \label{fig:Config7_PET}
\end{figure}

The 2D PET image reconstructed from the experimental data of i-TED-A and i-TED-B during the beam-off intervals for configuration (I) is displayed in panel (a) of Fig.~\ref{fig:Config7_PET}. The five irradiated layers are clearly observed and well resolved from each other. The maxima are well correlated with the position of the irradiated layers, as discussed below. The graphite layer is not visible in this image reconstruction because it was outside of the geometric PET field-of-view.

Fig.~\ref{fig:Config7_PET}-(b) shows the x-axis projection of the 2D PET distribution together with a MC simulation of the $\beta^{+}$ emission of the activated layers. The weight of each individual layer contribution was chosen to match the height of the experimental distribution. 
While the experimental distributions for the first, second and last sample slots are relatively well reproduced by the MC simulations, the third and fourth distribution positions are slightly broader than those reconstructed from the MC calculations. A plausible explanation for this effect might be that the PLA matrix did not fit these two layers perfectly, and thus the mean free path of the $\beta^{+}$ particles coming out from those irradiated layers was larger, thus leading to an additional broadening of the measured distributions at those locations.

The quality of the PET image can be quantified from the values listed in Tab.~\ref{tab:Config7_PET_Resolution}. The first column indicates the Nylon sample position. The second and third columns show the peak position from a Gaussian fit and shifts ($\Delta x$) with respect to the expected values. The width ($\sigma$) of the experimental distribution is listed in the 4$^{th}$ column, whereas the last column shows the widths ($\sigma$) from a Gaussian fit of the MC distribution.
\begin{table}[htb]
    \centering
    \begin{tabular}{c|c|c|c|c}\hline
    Nylon\# & x [mm] & $\Delta x$ [mm] & $\sigma_{Exp}$ [mm] & $\sigma_{MC}$ [mm] \\ \hline
    1 & 33.46(4) & -0.46(4) & 4.16(4) & 4.22\\ 
    2 & 14.33(3) & 0.67(3) & 4.41(3) & 4.03\\ 
    3 & -4.06(2) & -0.06(2) & 4.22(2) & 3.94\\ 
    4 & -23.11(3) & -0.89(3) & 5.53(4) & 4.17\\ 
    5 & -45.29(6) & -0.71(6)& 4.07(7) & 3.99\\ \hline    
    \end{tabular}
    \caption{The first column depicts the Nylon sample number for Configuration (I). The following columns show the values obtained for the mean value $x$ reconstructed for each layer, its deviation with respect to the expected quantity ($\Delta x$), the width of each experimental distribution $\sigma_{Exp}$ and the expected width $\sigma_{MC}$.}
    \label{tab:Config7_PET_Resolution}
\end{table}

There is a fair overall agreement between simulation and experiment. Sample positions are reconstructed via PET imaging always within a deviation of less than 1~mm. Variations in spatial resolutions are also, at most, of 1~mm with respect to expected MC values. 

These results for Configuration (I) allow one to anticipate information on the complementarity between Compton and PET imaging. Whereas the beam-on Compton image for 4.4~MeV $\gamma$-rays is a relatively broad distribution (FWHM$\sim$ 25~mm) but clearly correlated to the Bragg peak (Fig.~\ref{fig:Compton_Run7}), the beam-off PET technique offers a much finer resolution and sensitivity, although linked to the more indirect signature of the $\beta^{+}$ emitters.  

\subsection{Results for Configuration (II)}

\subsubsection{Compton imaging in the beam-on lapses}
\begin{figure}[!htbp]
    \centering
   \begin{tabular}{c c}
    \includegraphics[width=0.48\columnwidth]{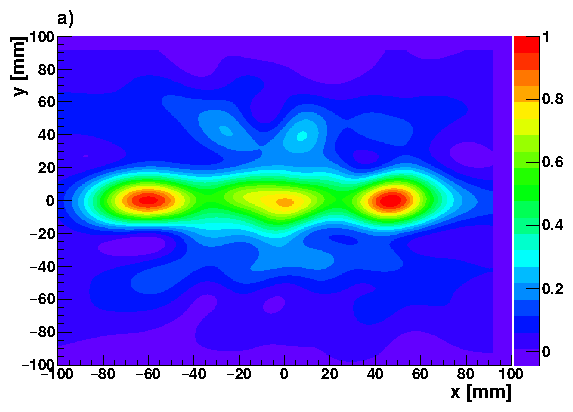} &
    \includegraphics[width=0.48\columnwidth]{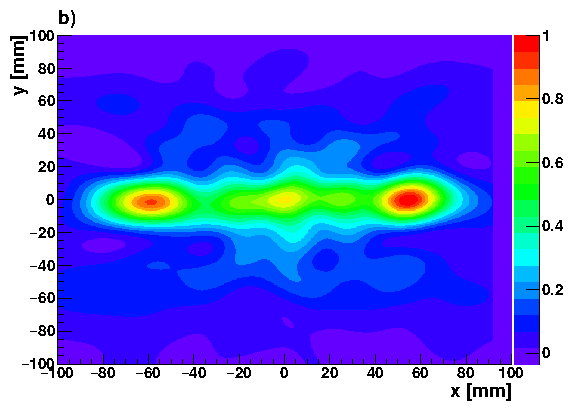} \\
    \includegraphics[width=0.48\columnwidth]{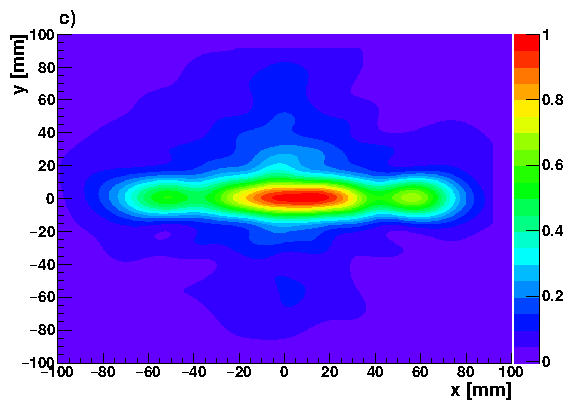} &
    \includegraphics[width=0.48\columnwidth]{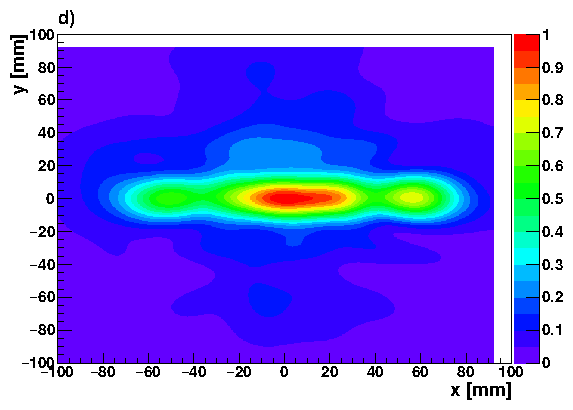} \\
    \end{tabular}
    \caption{Reconstructed experimental (top) and MC-calculated (bottom) Compton images from 4.4~MeV prompt $\gamma$-rays in configuration (II) for i-TED-A (left) and i-TED-B (right).}
    \label{fig:Compton_Run9}
\end{figure}

Fig~\ref{fig:Compton_Run9} shows the reconstructed Compton images for this configuration, both for the experimental data and the MC-simulation. 

The most relevant aspect of the 2D distribution reconstructed for this configuration are the two prominent maxima in the experimental images, with comparable strengths, separated by about 120~mm. The first maximum at positive x-values agrees well with the location of the proton energy beam degrader. The second maximum on the left-hand side (negative x-values) coincides with the one found in configuration (I), which is therefore consistent with the position of the thick graphite layer or beam stopper. In addition, in the central region of the experimental Compton image one can appreciate another maximum, with a peak value that is only $\sim$20\% lower than the two main lateral peaks.
Interestingly, for this configuration the reconstructed MC images do not reflect what is observed experimentally. The images, for both i-TED modules show a maximum in the central part of the image, where the Compton detection efficiency is maximum. It is worth to mention that between the central part and the positions where the thick graphite stopper and proton energy beam degrader are placed there is a difference of about $\sim$30\%. Almost the same value but in the opposite direction compared to the experimental images. At this moment we can only ascribe this notable discrepancy to possible deficiencies in the cross-sections present in the evaluated libraries used for the calculation, especially at low proton energies.  

\subsubsection{PET imaging in the beam-off lapses}
\begin{figure}[!htbp]
    \includegraphics[width=\columnwidth]{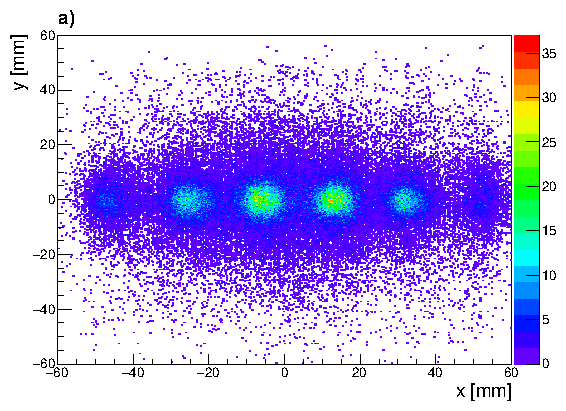} \\
    \includegraphics[width=\columnwidth]{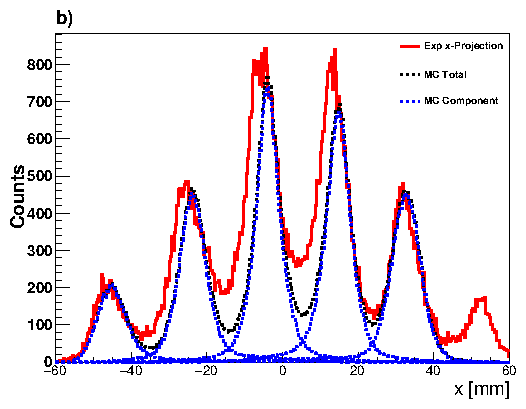} \\
    \caption{Experimental 2D PET image for Configuration II (a). PET image x-projection (red) together with the MC simulations of the individual images (blue) and the total MC distribution (black) are shown in panel (b).}
    \label{fig:PET_Config9}
\end{figure}

Fig.~\ref{fig:PET_Config9} shows the results obtained for the PET reconstruction of configuration (II).
Interestingly, in addition to the five Nylon layers, the signature from $\beta^{+}$ annihilation events at the energy degrader is well observed at the right-hand side of the image (x$=58$~mm), a position which is close to the edge of the PET field of view between both i-TED modules.  

According to Fig.~\ref{fig:PET_Config9}-(b) and the values reported in Tab.~\ref{tab:Config9_PET_Resolution} the positions reconstructed for the second and third layers are larger than those found in Configuration (I). The maximum deviation is of 1.85(4)~mm, although the average deviation for PET-reconstructed positions is still of 1.0(5)~mm.

\begin{table}[htb]
    \centering
    \begin{tabular}{c|c|c|c|c}\hline
    Nylon \# & x [mm] & $\Delta x$ [mm] & $\sigma_{Exp}$ [mm] & $\sigma_{MC}$ [mm] \\ \hline
    1 & 32.28(5) & 0.72(5) & 4.19(5) & 4.23 \\ 
    2 & 13.15(4) & 1.85(4) & 4.4(4) & 4.040 \\ 
    3 & -5.35(4) & 1.35(4) & 4.57(5) & 3.95 \\ 
    4 & -24.71(5) & 0.71(5) & 5.02(6) & 4.17 \\ 
    5 & -46.5(2) & 0.5(2) & 4.1(2) & 4.0 \\ \hline  
    \end{tabular}
    \caption{The first column depicts the Nylon sample number for configuration (II) starting from beam direction. The following columns show the values obtained for the mean value $x$ of each layer, its deviation with respect to the expected quantity ($\Delta x$), its with $\sigma_{Exp}$ and the expected width value $\sigma_{MC}$.}
    \label{tab:Config9_PET_Resolution}
\end{table}

Apart from confirming the general conclusions obtained for Configuration (I), the measurement for Configuration (II) also shows that MC simulations for evaluating the Compton-imaging performance are less reliable than the corresponding calculations for PET imaging. This is to be ascribed, on one side, to the better knowledge of the detector spatial response and overall performance for low-energy 511~keV $\gamma$-rays than for high-energy 4.4~MeV $\gamma$-rays and, on the other side, to the significant inaccuracies of ion-inelastic cross sections and $\gamma$-ray emission yields in the evaluated libraries used for MC studies, which may vary also significantly with ion-beam energy.



\subsection{Results for Configuration (III)}


\subsubsection{Compton imaging in the beam-on lapses}

\begin{figure}[!htbp]
    \centering
   \begin{tabular}{c c}
    \includegraphics[width=0.48\columnwidth]{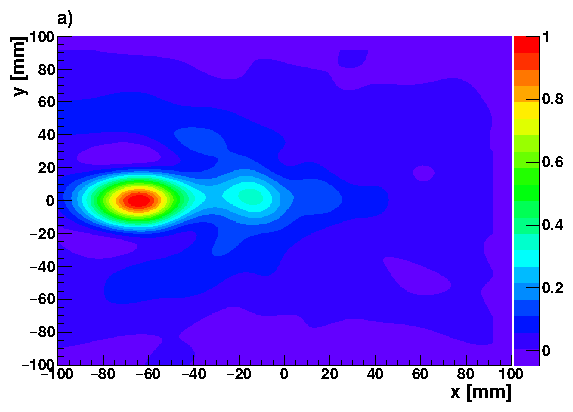} &
    \includegraphics[width=0.48\columnwidth]{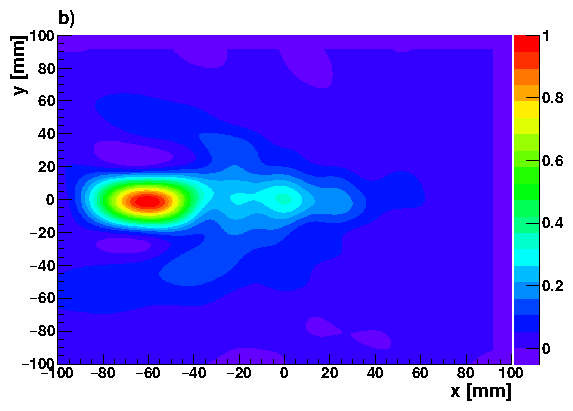} \\
    \includegraphics[width=0.48\columnwidth]{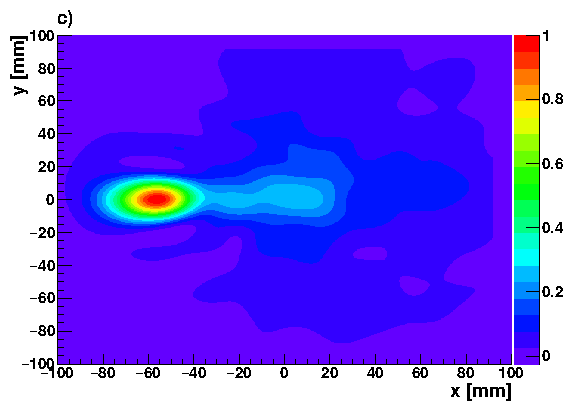} &
    \includegraphics[width=0.48\columnwidth]{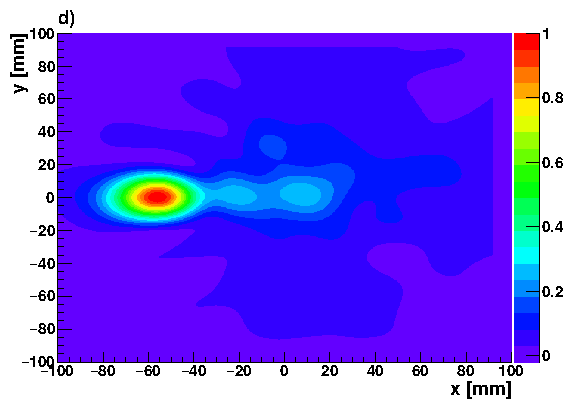} \\
    \end{tabular}
    \caption{Reconstructed experimental (top) and MC-calculated (bottom) Compton images from 4.4~MeV prompt $\gamma$-rays in configuration (III) for i-TED-A (left) and i-TED-B (right).}
    \label{fig:Compton_Run10}
\end{figure}

The results shown in Fig.~\ref{fig:Compton_Run10} for Configuration (III) are quite close to those found for Configuration (I) owing to similarities in the samples assembly, where essentially only the Nylon has been replaced by PMMA. These results help to assess the good repeatability of the Compton imaging technique and will not be discussed further. 

\subsubsection{PET imaging in the beam-off lapses}

\begin{figure}[!htbp]
    \includegraphics[width=\columnwidth]{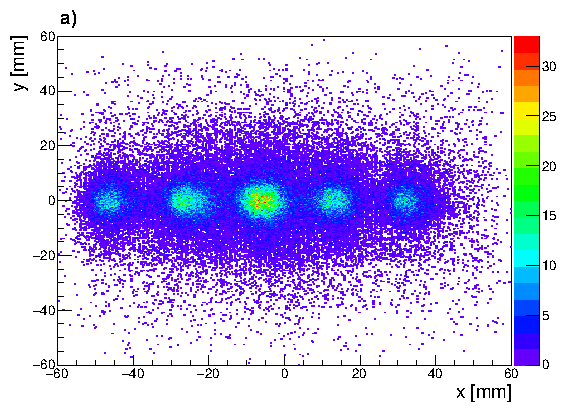} \\
    \includegraphics[width=\columnwidth]{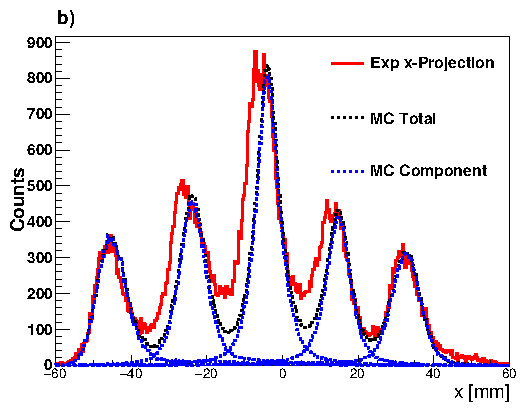} \\
    \caption{Experimental 2D PET image for Configuration (III) is shown in panel a). Panel b) shows the 1D PET image x-projection (red) together with the Monte Carlo simulations of the individual images (blue) and the total (black).}
    \label{fig:PET_Config10}
\end{figure}

The results obtained for both i-TED detectors in PET mode for Configuration (III) are illustrated in Fig.~\ref{fig:PET_Config10} and quantified in Tab.~\ref{tab:Config10_PET_Resolution}. As for the previous configurations, the five layers can be clearly resolved and located quite reliably at the true positions, with an average deviation of only 1.1(8)~mm. This result confirms therefore the high accuracy and the good repeatibility of the PET technique for this type of measurements.   




\begin{table}[htb]
    \centering
    \begin{tabular}{ c|c|c|c|c}\hline
    PMMA \# & x [mm] & $\Delta x$ [mm] & $\sigma_{Exp}$ [mm] & $\sigma_{MC}$ [mm]\\ \hline
    1 & 32.34(5) & 0.66(5) & 4.09(6) & 4.22 \\ 
    2 & 12.82(5) & 2.18(5) & 4.76(6) & 4.03\\ 
    3 & -5.67(3) & 1.67(5) & 4.39(4) & 3.91\\ 
    4 & -24.93(5) & 0.94(5) & 5.20(6) & 4.15\\ 
    5 & -46.1(1) & 0.1(1) & 4.0(1) & 4.0\\ \hline     
    \end{tabular}
    \caption{The first column indicates the PMMA sample number starting from beam direction. The following columns show the values obtained for the mean value $x$ of each layer, its deviation with respect to the expected quantity ($\Delta x$), its with $\sigma_{Exp}$ and the expected width value $\sigma_{MC}$.}
    \label{tab:Config10_PET_Resolution}
\end{table}

The results obtained for Configuration (III) confirm the conclusions obtained for the previous two configurations and, in particular, allow one to assess the good repeatibility of both Compton and PET techniques, given the good agreement with the results found in Configuration (I). In this respect, it is worth to emphasize that Configuration (III) was measured on a different day than Configuration (I), after samples were exchanged from Nylon to PMMA and detectors were shut-down and brought up again. Finally, to a certain extent, the discrepancies of 1-2~mm found for the PET location of layers 2, 3 and 4 between Configuration (I) and (III) could be also ascribed to experimental inaccuracies of the sample holder assembly.


\section{Discussion, conclusion and next steps}\label{sec:6}

In this work we have demonstrated the possibility to use two modular Compton cameras in a front-to-front and synchronous configuration in conjunction with a pulsed proton beam to perform in-situ prompt $\gamma$-ray Compton imaging and PET imaging, in a similar manner as conceptually proposed in Ref.~\cite{Parodi:16}. This pilot experiment comprised a variety of materials and configurations, thereby aiming at evaluating both imaging techniques in terms of accuracy, sensitivity, repeatibility and complementarity. 

For the specific experimental conditions of the measurements reported herein with a low-energy proton beam, the $\gamma$-ray imaging performance of PET overcomes that of Compton imaging in terms of spatial resolution (6~mm versus 24~mm) and detection sensitivity. The latter aspect is difficult to quantify in absolute terms, but visual inspection of the PET- and Compton-images clearly shows the capability of PET to visualize 0.8~mm thin plastic layers, whereas Compton-imaging is sensitive only to the bulk of $\gamma$-rays produced in the thick graphite layers, where a significant energy-loss of the proton-beam takes place. The very different performance in terms of sensitivity may not be the most relevant characteristic for ion-range monitoring, but it represents an aspect of potential interest when applying therapeutic beams to body regions where there can be a sudden change in tissue density due to air, bone, etc. As discussed in Sec.~\ref{sec:1}, this result for the higher spatial sensitivity of PET confirms its better suitability for monitoring physiological processes and tumour response during treatments with therapeutic beams.

The repeatability of the imaging technique may be more relevant for its application in ion-range monitoring, as discussed in Ref.\cite{Lerendegui:2022}. The repeatibility of our system for PET imaging can be evaluated from the $\Delta x$ differences between measured maxima and reference MC values for the sample layers of the different configurations, as reported in tables~\ref{tab:Config7_PET_Resolution}, \ref{tab:Config9_PET_Resolution} and \ref{tab:Config10_PET_Resolution}. Maximum deviations of about 2~mm were found, with average deviations typically within $\sim$1~mm. 
On the other hand, to estimate the repeatability in the reconstructed Compton images one can use the position of the maximum in each 2D-distribution measured with each i-TED imager for each configuration. These values are shown in Table~\ref{tab:Compton_repeatibility}. Deviations of up to 2.3~mm were found for Compton imaging, with an average value of 1.2(9)~mm for the three configurations. This result is comparable to the average variations measured for the samples with the PET technique. 

\begin{table}[!htbp]
    \centering
    \begin{tabular}{c| c c c}\hline
     &   &  Configuration & \\
     &  &   Max. Position in Compton (mm) &  \\  
        i-TED &  (I) &  (II) & (III) \\ \hline
        A & -71.17  & -63.50 & -70.50 \\ 
        B & -68.8  & -64.16 & -71.16 \\ \hline
        $\Delta x$ & 2.3  & 0.66 & 0.66 \\ \hline
    \end{tabular}
    \caption{Maximum emission position for the thick graphite beam-stopper determined via Compton imaging for the three configurations and both i-TED modules. The bottom raw shows the differences for each configuration between maxima of both imagers i-TED-A and i-TED-B.}
    \label{tab:Compton_repeatibility}
\end{table}

It is worth to recall, however, that the Compton technique, despite its limitations in image resolution and detection sensitivity, it shows the inherent advantage of being sensitive to the high-energy $\gamma$-rays that are both spatially and temporally directly correlated to the Bragg peak, thus providing a more direct pristine information for ion-range monitoring.

Apart from the intrinsic systematic advantages and drawbacks of each imaging technique, an aspect which is of pivotal relevance for achieving real-time ion-range monitoring is the statistical significance attainable with each technique in a short period of time. In-room PET scanners have a much higher efficiency ($\varepsilon^{PET}\sim 2\times 10^{-2}$) than prompt $\gamma$-ray imagers based on mechanical collimation ($\varepsilon^{Slit}\leq 10^{-3}$)~\cite{Krimmer:18} used for clinical studies\cite{Smeets:2012}. However, for the combined system reported here based on i-TED modules, the efficiency for prompt $\gamma$-rays is much larger than that of a slit Camera, owing to the electronic collimation technique and the large intrinsic and geometric efficiency of the position-sensitive detectors implemented~\cite{Lerendegui:2022}. For the exploratory measurements carried out in this work the average count rates in-spill and off-spill for each configuration and technique, Compton or PET, are reported in table~\ref{tab:count_rates}.
\begin{table}[!htbp]
    \centering
    \begin{tabular}{c| c c c}\hline
     &   &  Configuration & \\
     &  &   Avg. Count Rate (Hz) &  \\  
         &  (I) &  (II) & (III) \\ \hline
        Compton in-spill & 13324  & 6642 & 14975 \\ 
        PET off-spill & 61  & 29 & 20 \\ \hline
    \end{tabular}
    \caption{Average count rates for Compton imaging in-spill and for PET imaging off-spill.}
    \label{tab:count_rates}
\end{table}
Thus, on average, a factor 300 higher counting rate was registered in-spill for Compton imaging, when compared to the statistics off-spill with the PET technique. This result obviously depends on the particular duty-cycle of the proton beam and, therefore, needs to be re-evaluated with clinical beam conditions in future experiments. For the particular case that a similar a counting-rate ratio of 300:1 would be attained in a clinical treatment, in order to achieve a similar counting statistics with both imaging methods a pulsed beam with a duty-cycle of 0.01, or smaller, would become convenient from the point-of-view of real-time range monitoring. Furthermore, it would allow one to add a similar level of statistics from both imaging techniques, while the systematic uncertainties could be kept under better control thanks to the inherent differences of each imaging approach.

In summary, to a large extent both PET and PG Compton techniques may be rather complementary and they could be combined in a synergetic fashion to improve ion-range monitoring both in terms of statistical and systematic accuracy. Therefore, the possibility to combine them in a single and dedicated system, similar to the one reported here, seems a reasonable step forward that is worth exploring further in this field. 

In future works we aim at overcoming the main limitation of the present study, which was the low proton-beam energy of 18~MeV, significantly smaller than common clinical values (100-200~MeV). For this reason the beam on/off lapses used in this work had to be significantly large, of several 100~s and a special target-phantom with several layers had to be employed. Next steps to develop further the hybrid technique comprise the realization of measurements using average clinical beam conditions and conventional water and PMMA phantoms. Using a pulsed beam of a high energy with a large phantom and a suitable duty cycle, it may become possible to correlate sequential PG-Compton and PET-images in- and off-spill, respectively, as proposed also in Ref.~\cite{Parodi:16}. In this respect, there are two main aspects to research in the future experiment. First, the performance of the detection system at the high counting rate conditions of the high-energy and high-intensity clinical beams needs to be validated. Second, one has to determine which is the optimal duty-cycle for the proton beam to accomplish both PG Compton and PET imaging for a proper trade-off of statistical and systematic accuracy. The aforediscussed pulsed-beam duty-cycle requirement of less than 0.01 aligns very well with the application of this new methodology in treatments with superconducting synchrocyclotrons, which produce a pulsed beam (few $\mu$s, every 1–2 ms)~\cite{Yap21}, and also with recent developments related to hybrid delivery approaches in flash therapy~\cite{Jolly20}. 

\section*{Acknowledgment}
This work has been carried out in the framework of a project funded by the European Research Council (ERC) under the European Union's Horizon 2020 research and innovation programme (ERC Consolidator Grant project HYMNS, with grant agreement No.~681740). The authors acknowledge support from the Spanish Ministerio de Ciencia e Innovaci\'on under grants PID2019-104714GB-C21, FPA2017-83946-C2-1-P, FIS2015-71688-ERC, FPA2016-77689-C2-1-R, RTI2018-098117-B-C21, CSIC for funding PIE-201750I26 and from the European H2020-847552 (SANDA). This work was partially supported by Generalitat Valenciana PROMETEO/2019/007. T. Rodriguez-Gonzalez acknowledges the Spanish FPI predoctoral grant. We would like to thank the crew at the Electronics Laboratory of IFIC, in particular Manuel Lopez Redondo and Jorge N\'acher Ar\'andiga for their excellent and efficient work.

\section*{Author contributions}
JB-C Investigation, methodology, formal
analysis, data curation, visualization, writing-original draft. JL-M investigation, methodology, formal
analysis, data curation. IL hardware, software, visualization. CG conceptualization, methodology,
supervision, investigation, formal analysis. TR-G  Investigation, methodology, formal
analysis, data curation. MCJ-R conceptualization, methodology, supervision, investigation, formal analysis. CD-P conceptualization, methodology, supervision, writing-review \& editing, funding acquisition, investigation, formal analysis.

\section*{Conflict of interest}
The authors declare that they have no known competing financial interests or personal relationships that could have appeared to influence the work reported in this paper.

\bibliography{Bibliography}

\end{document}